\documentclass[twocolumn,preprint,showpacs,preprintnumbers,amsmath,amssymb,showkeys]{revtex4_cgh}

\usepackage{graphicx}% Include figure files
\usepackage{dcolumn}% Align table columns on decimal point
\usepackage{bm}% bold math
\usepackage[center]{subfigure}

\begin{document}

\preprint{APS/123-QED}

\title{Screening effects on field emission from arrays of (5,5) carbon nanotubes: Quantum-mechanical simulation}

\author{Guihua Chen, Weiliang Wang, Jie Peng, Chunshan He,\\ Shaozhi Deng, Ningsheng Xu}
\author{Zhibing Li}
 \email{stslzb@mail.sysu.edu.cn }
 \affiliation{State Key Laboratory of Optoelectronic Materials and
 Technologies, School of Physics and Engineering, Sun Yat-Sen University, Guangzhou 510275,
China }

\date{\today}% It is always \today, today,

\begin{abstract}
The simulation of field electron emission from arrays of micrometer-long
open-ended (5, 5) carbon nanotubes is performed in the framework of quantum
theory of many electrons. It is found that the applied external field is
strongly screened when the spacing distance is shorter than the length of the
carbon nanotubes. The optimal spacing distance is two to three times of the
nanotube length, slightly depending on the applied external fields. The
electric screening can be described by a factor that is a exponential function
of the ratio of the spacing distance to the length of the carbon nanotubes. For
a given length, the field enhancement factor decreases sharply as the screening
factor larger than 0.05. The simulation implies that the thickness of the array
should be larger than a value but it does not help the emission much by
increasing the thickness a great deal.
\end{abstract}

\pacs{73.22.-f  73.21.-b  79.70.+q}% PACS, the Physics and Astronomy
\keywords{field emission, carbon nanotube, screening effect }%Use showkeys class option if keyword
\maketitle

\section{\label{sec:level} Introduction}

For the application of carbon nanotubes (CNTs) as the cold cathodes of field
electron emission (FE), the ideal structure is to arrange the CNTs into an
aligned array. It has been observed that the spacing distance of CNTs affects
FE properties remarkably.\cite{a1,a2,a3,a4,a5,a6,a7,a8,a9} It is clear that
denser and longer CNT bundle has stronger screening effect that would reduce
the field electron emission of each CNT. On the other hand, larger spacing
distance decreases the number of CNTs in a unit area, which leads to weaker
mean emission current density. It is important to describe the screening
quantitatively and to find out its effect on the FE ability. The calculation of
Nilsson \textit{et al.}\cite{a1} suggested that the optimum spacing distance
would be twice the length of the CNTs. However, Jung Sang Suh \textit{et
al.}\cite{a2} showed that the emission current density is optimized when the
length of the CNTs is equal to the spacing distance. By solving Laplace's
equation, Bocharov and Eletskii\cite{a9} found that the emission current
density has the maximum value when the spacing distance is half of the length
of the CNTs. More careful studies on this topic would obviously be useful. In
this paper, we will simulate the FE of the array of single-walled carbon
nanotubes (SWCNTs) with a quantum/molecular hybrid method.

Under the applied external field (denoted by $F_{appl}$), it has been known
that the SWCNT is charged.\cite{a10} Therefore, the SWCNTs of the array are
coupled to each other through the Coulomb interaction. The excess charges are
also the origin of the screening. The competition between the screening effect
and the density of CNTs would be complicated by the field enhancement factor
that is length-dependent. The field enhancement factor was presumably
proportional to the aspect ratio of the SWCNT. However, recent quantum
simulations revealed that the field penetration at the apex of the SWCNT is
significant.\cite{a10,a11} The field penetration depends on the value of the
applied external field, thereby it is also related to the screening effect. The
enhancement factor of the array has been calculated by the classical
method.\cite{a9} When the charge redistribution and the screening effect are
taken into account, as will be shown in the present paper, the field
enhancement factor of the array is different from the classical one obviously.

To estimate the excess charge distribution and the field penetration at the
apexes of SWCNT arrays, it requires a large-scale simulation that should
reflect both the quantum electron structure at the apex and the Coulomb
interaction over the tubes. In experiments, the length of CNTs is usually in
micrometers, while the radius is in nanometers. A huge number of freedoms are
involved. For instance, the (5, 5) type SWCNT of 1\textrm{ $\mu$m} length
consists of about $10^5$ carbon atoms. Limited by the computational efficiency
and resources, all ab initio studies so far can only simulate the local
properties involving hundreds of carbon atoms. As the electronic properties are
sensitive to both the detailed atomic arrangement (i.e., the location of
defects, adsorbates, and the chirality) and the distribution of excess charges
over the whole tube, it is a big challenge to simulate a SWCNT array that
consists of SWCNTs with length in the order of micrometer. Only recently has it
been possible to tackle an individual SWCNT of realistic size in the FE
conditions by a multiscale method involving quantum mechanics and molecular
mechanics.\cite{a10,a11} In the present paper, we adopted this method to
simulate the FE of the SWCNT array, of which the length is in the micrometer
scale. The structure of the SWCNTs is specified to the (5, 5) armchair type.
The dangling bonds in the open mouths of the SWCNTs are saturated by hydrogen
atoms. We further assume that the SWCNTs are vertically mounted on a metal
surfaced uniformly. Therefore only one SWCNT is required to be dealt with in
the environment of other SWCNTs. Each SWCNT of the array is treated as a
replica of the SWCNT under simulation.

In Section II, the simulation method is reviewed briefly. The simulation
results are presented and discussed in Section III. The dependence of the
screening on the spacing distance, the length, and the applied external fields
are discussed in Section IV, where a factor is introduced to describe the
screening, and the correlation between this factor and the field enhancement
factor is presented. The last section gives the conclusions.

\section{\label{sec:level} Simulation Method}

The CNT array for FE is a typical multiscale system. In our model, the CNT
array is under the uniform applied external field, whose direction is parallel
to the axes of the CNTs, and the SWCNTs are mounted vertically and formed a
regular square lattice on the cathode surface. The distance of two nearest
neighbored SWCNTs (to be referred to as "spacing distance") and the length of
SWCNTs in the array are denoted by \textit{d} and \textit{L}, respectively.
When a field is applied to the SWCNT array, the electrons have opportunity to
emit into vacuum through the apexes of the SWCNTs by quantum tunneling.

%\begin{figure}
%\includegraphics[scale=0.5]{fig_1}
%\caption{\label{fig_1} The schematic setup of field electron emission of a
%SWCNT array. The arrow lines represent the applied external field.}
%\end{figure}

Since electrons are emitted from the apex of each SWCNT by quantum tunneling,
the apex part must be treated by quantum mechanics. The part on the substrate
side mainly affects the field emission through Coulomb potential of the excess
charges, so  it can be treated by a semiclassical method.\cite{a10,a11}
Therefore we should divide each SWCNT into a quantum region and a semiclassical
region. The quantum region is dealt with on atomic scale where the density
matrix of the electrons is obtained quantum mechanically. The quantum region
should be large enough to ensure that the artificial division does not affect
the physical results seriously. By our experience, the proper size of the
quantum region is much bigger than that the standard ab initio methods could
deal with. We have to further divide the quantum region into sub-regions. Each
sub-region together with its adjacent sub-regions forms a subsystem that is
dealt with by the modified neglect of diatomic overlap (MNDO)\cite{Dewar}
semiempirical quantum mechanical method (here the MOPAC software has been
used). The excess charges outside the subsystem being dealt with are treated as
point charges. Their contribution to the subsystem being dealt with is through
the Coulomb interaction. To accelerate the simulation, we first simulate an
isolated subsystem under various external fields. Assuming all subsystems
except that contains the apex are resemble to the isolated subsystem, their
electron density can be read out from the database, which has been constructed
by simulating the isolated subsystem, with the entry of electric field that is
the superposition of the applied external field and the fields contributed by
the excess charges in other parts of the array as well as by the image charges.
With this acceleration algorithm, we are able to deal with a quantum region of
length over 900\textrm{ nm}.

Although in principle we can deal with an entire isolated SWCNT quantum
mechanically, the connection of the SWCNT and the cathode is too complicate for
a full quantum treatment. It is convenient to treat the part of SWCNT
connecting the cathode as the semiclassical region. In the semiclassical
region, the Coulomb potential is governed by Poisson's equation. The boundary
condition of the metal surface is guaranteed by the image charges of the excess
charges of the SWCNTs. It should be noted that even in the semiclassical region
the electron energy band structure originating from the quantum mechanics
should be taken into account. For the (5, 5) SWCNT, there are both experimental
and theoretical evidences for the constant density of state (DOS) in the
vicinity of the neutrality level.\cite{a12,a13,a14,a15} It turns out that the
excess charge density can be approximated by a linear function of the
longitudinal coordinate of the SWCNT.\cite{a16}

The coupling of quantum region and semiclassical region is through the
quasithermodynamic equilibrium condition which assumes that the chemical
potential (Fermi level) is a constant over the entire array. To avoid the
complexity arose from the Schottky junction that would emerge at the back
contact in principle, we simply assume the Fermi level of the SWCNTs is
5.0\textrm{ eV} below the vacuum potential in the absence of applied external
field. The density of the excess charge ("excess density" for simplicity)
calculated separately in quantum and semiclassical regions should coincide at
an overlap place of two regions. The self-consistent excess density of the
entire SWCNT array is achieved through iterations that contain a small loop and
a big loop. In the small loop, the subregions of the quantum region are dealt
with one by one, and repeated until that a converged electron density in the
quantum region is obtained. In the big loop, the quantum region and the
semiclassical region are dealt with alternatively until the self-consistent
charge distribution is achieved.

\section{\label{sec:level} Simulation Results}

It has been shown that the vacuum potential barrier in the circumambience of
the SWCNT is high and thick\cite{a10}, therefore the electrons would most
probably emit forward from the first layer of the tip. The transmission
coefficient (\textit{D}) can be estimated by the WKB approximation

\begin{equation}\label{eq_1}
D =
\exp\Bigl[-\frac{2}{\hbar}\int\sqrt{2m[U(z)-E_\textrm{f}]}\mathrm{d}z\Bigr],
\end{equation}

\noindent where \textit{U(z)} is the electron energy potential, $E_\textrm{f}$
is the Fermi energy, and the integral is over the classical forbidden region
where $U(z)-E_\textrm{f}>0$. We have assumed that the electrons possess the
Fermi energy.

With \textit{D} in hand, the emission current of each SWCNT is estimated by

\begin{equation}\label{eq_2}
I = \nu q_{\textrm{exc}}D,
\end{equation}

\noindent where $q_{\textrm{exc}}$ are the extra electrons of the first layer
atoms, and $\nu$ is the collision frequency (the number of electrons hitting
the barrier per unit time) that can be estimated from the average kinetic
energy of $\pi^\textrm{*}$ electrons as $E_\textrm{k}(\pi^\textrm{*}) / h$,
which is approximately equal to $10^{14}$ Hz.

We should focus at the electrostatic potential \textit{U(z)} in the vicinity of
the apex of the SWCNT, which determines the major feature of the FE of the
array.

\subsection{\label{sec:level2} Varying spacing distance}

\begin{figure}
\includegraphics[scale=0.6]{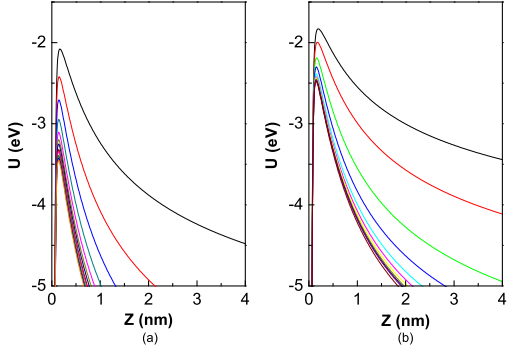}
\caption{\label{Pvsd} (Color online) The potential of different spacing
distances (\textit{d}) in $F_{\textrm{appl}}$=12.0\textrm{ V/$\mu$m}. (a)
\textit{L}=1.00\textrm{ $\mu$m}. The curves from up to down are corresponding
to different \textit{d} ranged from 0.50\textrm{ $\mu$m} to 2.50\textrm{
$\mu$m} with step of 0.25\textrm{ $\mu$m}. The potential of
\textit{d}=4.00\textrm{ $\mu$m} and that of the individual SWCNT are also
plotted, but they are invisible since they coincide with the lowest curve. (b)
\textit{L}=0.75\textrm{ $\mu$m}. The curves from up to down are corresponding
to different \textit{d} ranged from 0.50\textrm{ $\mu$m} to 2.50\textrm{
$\mu$m} with step of 0.25\textrm{ $\mu$m}. The potential of
\textit{d}=0.375\textrm{ $\mu$m} and that of the individual SWCNT are also
plotted, and corresponding to the uppermost and lowermost curves.}
\end{figure}

In Fig. \ref{Pvsd}(a)/(b), we plot \textit{U(z)} of different spacing distances
under the applied external field, $F_{\textrm{appl}}$=12.0\textrm{ V/$\mu$m}.
The length of the tubes, i.e., the thickness of the array, is 1.00\textrm{
$\mu$m} in Fig. \ref{Pvsd}(a) and 0.75\textrm{ $\mu$m} in Fig. \ref{Pvsd}(b).
The \textit{Z} axis has its origin at the last atom of the SWCNT and is
parallel to the direction of the tube axis. The Fermi level of the electrons in
the nanotubes is assumed to be -5.0\textrm{ eV} in our simulation. Therefore
the potential profiles shown in Fig. \ref{Pvsd} are in fact the apex-vacuum
potential barriers that control the probability of electron tunneling. The
observation of the barrier height lowering as \textit{d} increasing is
consistent with the FE mechanism of field-reduced barrier\cite{a10} since
larger \textit{d} should lead to smaller field screening and increase the
effective field applying to the SWCNTs. The sharper shape of Fig. \ref{Pvsd}(a)
comparing with Fig. \ref{Pvsd}(b) is a consequence of the field enhancement.
The curves almost coincide with each other for \textit{d}$>$1.5\textrm{ $\mu$m}
in Fig. \ref{Pvsd}(a)/(b), implying that the screening effect is negligible
when \textit{d}$>$1.5\textrm{ $\mu$m}.

\begin{figure} %并排插入两个子图形
\centering \subfigure[]{
\label{Jvsd_a} %%标记第一个子图形
\includegraphics[scale=0.6]{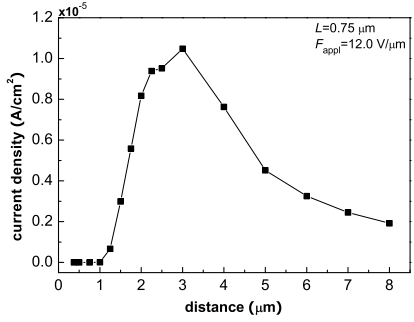}}
\hspace{0.1in} \subfigure[]{
\label{Jvsd_b} %%标记第二个子图形
\includegraphics[scale=0.6]{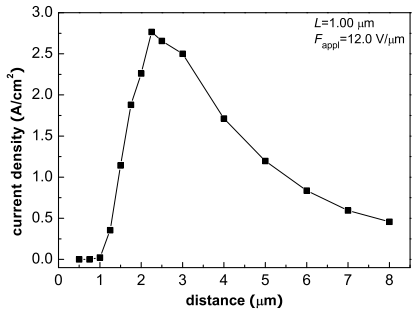}}
\hspace{0.1in} \subfigure[]{
\label{Jvsd_c} %%标记第二个子图形
\includegraphics[scale=0.6]{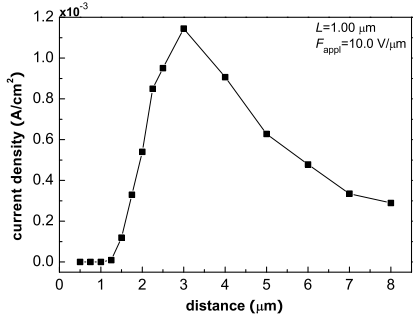}}
\caption{The current density of the SWCNT array versus the spacing distance. (a) \textit{L}=0.75\textrm{ $\mu$m} and
$F_{\textrm{appl}}$=12.0\textrm{ V/$\mu$m}. (b) \textit{L}=1.00\textrm{ $\mu$m} and $F_{\textrm{appl}}$=12.0\textrm{
V/$\mu$m}. (c) \textit{L}=1.00\textrm{ $\mu$m} and $F_{\textrm{appl}}$=10.0\textrm{ V/$\mu$m}.}
\label{Jvsd} %%标记整个图形
\end{figure}

The mean current density from the array ("current density" for simplicity and
denoted by \textit{J}) against \textit{d} are presented in Fig. \ref{Jvsd} for
three sets of parameters. The lengths of SWCNTs are 0.75\textrm{ $\mu$m},
1.00\textrm{ $\mu$m}, and 1.00\textrm{ $\mu$m}; the external fields are
12.0\textrm{ V/$\mu$m}, 12.0\textrm{ V/$\mu$m}, and 10.0\textrm{ V/$\mu$m} for
the Figs. \ref{Jvsd}(a), (b), and (c), respectively. The current density is
very sensitive to both \textit{L} and $F_{\textrm{appl}}$ (the \textit{J} axes
for Figs. \ref{Jvsd}(a), (b), and (c) are multiplied by the factors $10^{-5}$,
$10^0$, and $10^{-3}$, respectively). It is notable that the emission is
turned-on at certain spacing distance, which is about 1.0\textrm{ $\mu$m} here,
roughly equal to the length of the tube. Comparing Figs. \ref{Jvsd}(a) , (b),
and (c), one may see that both the turn-on spacing distance and the maximum of
\textit{J} would depend on $F_{\textrm{appl}}$.

The parabolic decrease of the current density as the spacing distance getting
large can be easily understood by the fact that the current density is
proportional to the number of SWCNTs in a unit of area when the screening
effect is negligible.

\subsection{\label{sec:level2} Varying length}

\begin{figure}
\includegraphics[scale=0.6]{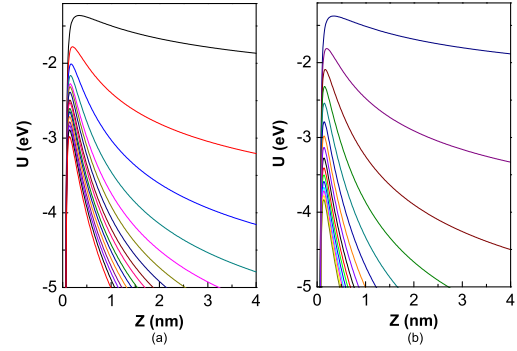}
\caption{\label{PvsL} (Color online) \textit{U(z)} for different \textit{L}
under the applied external field of 10.0\textrm{ V/$\mu$m}. The curves from up
to down are corresponding to different \textit{L} ranged from 0.25\textrm{
$\mu$m} to 4.00\textrm{ $\mu$m} with step of 0.25\textrm{ $\mu$m}. (a)
\textit{d}=0.75\textrm{ $\mu$m}. (b) \textit{d}=1.00\textrm{ $\mu$m}.}
\end{figure}

We plot the \textit{U(z)} for different \textit{L} under the applied external
field $F_{\textrm{appl}}$=10.0\textrm{ V/$\mu$m}, with \textit{d}=0.75 in Fig.
\ref{PvsL}(a) and 1.00\textrm{ $\mu$m} in Fig. \ref{PvsL}(b). The lowering of
apex-vacuum barrier is associated with the excess charge accumulation at the
apex. For the same \textit{L}, the barrier of larger \textit{d} is lower than
that of shorter \textit{d}. It implies that the array of larger \textit{d} can
accommodate more excess charges at each apex.

%\begin{figure} %并排插入两个子图形
%\centering \subfigure[]{
%\label{fig_4_a} %%标记第一个子图形
%\includegraphics[scale=0.5]{fig_4_a}}
%\hspace{0.1in} \subfigure[]{
%\label{fig_4_b} %%标记第二个子图形
%\includegraphics[scale=0.5]{fig_4_b}}
%\caption{\textit{U(z)} for different \textit{L} under the applied external
%field of 10.0\textrm{ V/$\mu$m}. The curves from up to down are corresponding
%to different \textit{L} ranged from 0.25\textrm{ $\mu$m} to 4.00\textrm{
%$\mu$m} with step 0.25\textrm{ $\mu$m}. (a) \textit{d}=0.75\textrm{ $\mu$m} ,
%(b) \textit{d}=1.00\textrm{ $\mu$m}.}
%\label{fig_4} %%标记整个图形
%\end{figure}

\begin{figure}
\includegraphics[scale=0.6]{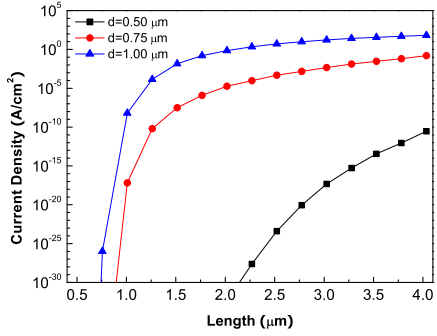}
\caption{\label{JvsL} (Color online) The current density versus the length of
the SWCNTs. The squares, circles, and triangles are corresponding to
\textit{d}=0.50\textrm{ $\mu$m}, \textit{d}=0.75\textrm{ $\mu$m}, and
\textit{d}=1.00\textrm{ $\mu$m}, respectively.}
\end{figure}

The current density as a function of \textit{L} is shown in Fig. \ref{JvsL}.
The squares, circles, and triangles in this figure are corresponding to the
data of \textit{d}=0.50\textrm{ $\mu$m}, \textit{d}=0.75\textrm{ $\mu$m}, and
\textit{d}=1.00\textrm{ $\mu$m}, respectively. From this figure, one sees that
the emission current densities of the array of short SWCNTs are negligible. The
minimum length for a significant current density (say $10^{-10}$\textrm{
A/$\textrm{cm}^2$}) is enlarged as the spacing distance decreases. The rapid
increase of the current densities is the consequence of the field enhancement.
However, when \textit{L} is large, the screening has significant effects and
the current densities only increase moderately. The array with
\textit{L}=0.5\textrm{ $\mu$m} can hardly emit electrons.

\subsection{\label{sec:level2} Varying applied external field}

\begin{figure}
\includegraphics[scale=0.6]{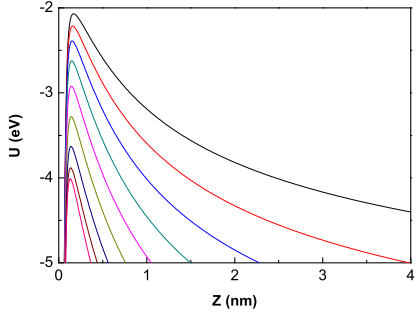}
\caption{\label{PvsF} (Color online) \textit{U(z)} for the $F_{\textrm{appl}}$
ranged from 7 to 15\textrm{ V/$\mu$m} (up to down) with \textit{d}=2.00\textrm{
$\mu$m} and \textit{L}=1.00\textrm{ $\mu$m}.}
\end{figure}

We have considered the array with \textit{d}=2.00\textrm{ $\mu$m} and
\textit{L}=1.00\textrm{ $\mu$m} for various applied external fields. The
electron potential \textit{U(z)} is presented in Fig. \ref{PvsF} for
$F_{\textrm{appl}}$ ranged from 7 to 15\textrm{ V/$\mu$m}.

%\begin{figure}
%\includegraphics[scale=0.5]{fig_6}
%\caption{\label{fig_6} \textit{U(z)} for the $F_{\textrm{appl}}$ ranged from 7
%to 15\textrm{ V/$\mu$m} (up to down) with \textit{d}=2.00\textrm{ $\mu$m} and
%\textit{L}=1.00\textrm{ $\mu$m}.}
%\end{figure}

\begin{figure}
\includegraphics[scale=0.6]{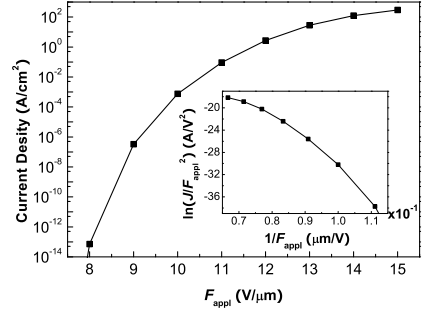}
\caption{\label{fig_4} The current density versus the applied external field
for the SWCNT array with \textit{d}=2.00\textrm{ $\mu$m} and
\textit{L}=1.00\textrm{ $\mu$m}. The insert shows the corresponding
Fowler-Nordheim plot.}
\end{figure}

The \textit{J}-$F_{\textrm{appl}}$ characteristic is given in Fig. \ref{fig_4}.
The inset of Fig. \ref{fig_4} is the Fowler-Nordheim plot. The nonlinear
Fowler-Nordheim plot implies that the mechanism for the FE of the SWCNT array
could be different from that of the metal plane emitters.

\section{\label{sec:level} Screening Factor}

\begin{figure}
\includegraphics[scale=0.6]{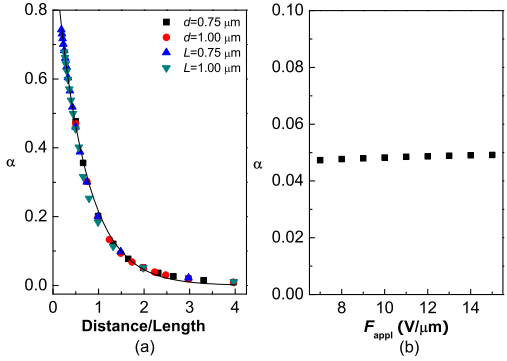}
\caption{\label{fig_5} (Color online) (a) The screening factor $\alpha$ versus
the ratio of the spacing distance to the length of the SWCNTs (\textit{d/L}).
(b) The screening factor $\alpha$ versus $F_{\textrm{appl}}$.}
\end{figure}

In order to describe the screening effect quantitatively, we define a screening
factor $\alpha$ as

\begin{equation}\label{eq_3}
\alpha = 1-\frac{V}{LF_{\textrm{appl}}},
\end{equation}

\noindent where \textit{V} is the voltage drop (related to the substrate) at
the middle point of the line connecting two neighborhood apexes. It should be
zero if there is no screening and equal to 1 if the array, as an ideal metal
layer of thickness \textit{L}, screens the field completely. The screening
factor $\alpha$ versus \textit{d/L} is shown in Fig. \ref{fig_5}(a) for four
sets of parameters: (1) \textit{d}=0.75\textrm{ $\mu$m},
$F_{\textrm{appl}}$=12.0\textrm{ V/$\mu$m}; (2) \textit{d}=1.00\textrm{
$\mu$m}, $F_{\textrm{appl}}$=12.0\textrm{ V/$\mu$m}; (3)
\textit{L}=0.75\textrm{ $\mu$m}, $F_{\textrm{appl}}$=10.0\textrm{ V/$\mu$m};
(4) \textit{L}=1.00\textrm{ $\mu$m}, $F_{\textrm{appl}}$=10.0\textrm{
V/$\mu$m}. The upper (lower) triangles are corresponding to various \textit{d}
with fixed \textit{L}=0.75 (1.00)\textrm{ $\mu$m}. The squares (circles) are
corresponding to various \textit{L} with fixed \textit{d}=0.75 (1.00)\textrm{
$\mu$m}. Notably, all points fall into a curve, implying that the screening
factor is a function of \textit{d/L}. The best fitting to the simulation (solid
curve) is

\begin{equation}\label{eq_4}
\alpha = \exp{(-d/0.65L)},
\end{equation}

For the array of \textit{d}=2.00\textrm{ $\mu$m} and \textit{L}=1.00\textrm{
$\mu$m}, the screening factor $\alpha$ is calculated with different
$F_{\textrm{appl}}$. As shown in Fig. \ref{fig_5}(b), we find that it is almost
independent of the applied external field. It implies that the screening factor
fixed by the ratio \textit{d/L} is an intrinsic feature of the array.

%\begin{figure}
%\includegraphics[scale=0.5]{fig_9}
%\caption{\label{fig_9} The screening factor versus the applied external field
%for \textit{d}=2.00\textrm{ $\mu$m} and \textit{L}=1.00\textrm{ $\mu$m}.}
%\end{figure}

\begin{figure}
\includegraphics[scale=0.6]{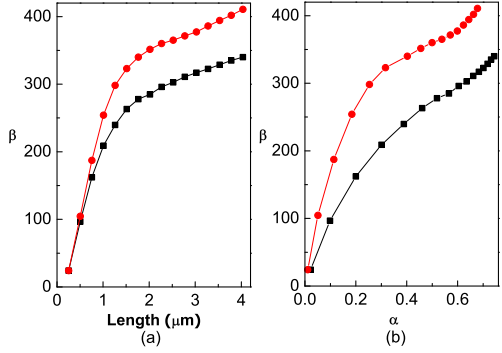}
\caption{\label{fig_6} (Color online) The field enhancement factor($\beta$) in
$F_{\textrm{appl}}$=10.0\textrm{ V/$\mu$m}. The squares are for
\textit{d}=0.75\textrm{ $\mu$m}, and the triangles are for
\textit{d}=1.00\textrm{ $\mu$m}. (a) $\beta$ versus the length of the SWCNTs;
(b) $\beta$ versus the screening factor.}
\end{figure}

From the data of \textit{U(z)}, we can estimated the local field strength
($F_\textrm{m}$) as the value of the slope of the barrier potential
\textit{U(z)} at the point of steepest decrease. An effective field enhancement
factor $\beta$ can then be defined as $F_\textrm{m}$/$F_{\textrm{appl}}$.
Figure \ref{fig_6} shows the correlation between $\beta$ and the tube length
(a), and the correlation between $\beta$ and the screening factor (b), for
fixed $F_{\textrm{appl}}$ and \textit{d} . The field enhancement factor of the
array is not proportional to the tube length, which is completely different
from an individual nanotube (or a metal tip). The increase of tube length
strengthens $\beta$ and the screening tends to reduce $\beta$ . But the
increase \textit{L} will lead to increase of screening. Therefore, the
dependence of $\beta$ on \textit{L} and \textit{$\alpha$} is complicate. If
there were no screening, $\beta$ in Fig. \ref{fig_6}(b) should increase faster.

%\begin{figure} %并排插入两个子图形
%\centering \subfigure[]{
%\label{fig_10_a} %%标记第一个子图形
%\includegraphics[scale=0.5]{fig_10_a}}
%\hspace{0.1in} \subfigure[]{
%\label{fig_10_b} %%标记第二个子图形
%\includegraphics[scale=0.5]{fig_10_b}}
%\caption{The field enhancement factor($\beta$) in
%$F_{\textrm{appl}}$=10.0\textrm{ V/$\mu$m}. The squares are for
%\textit{d}=0.75\textrm{ $\mu$m}, the triangles are for \textit{d}=1.00\textrm{
%$\mu$m}. (a) $\beta$ versus \textit{L}; (b) $\beta$ versus the screening factor
%($\alpha$).}
%\label{fig_10} %%标记整个图形
%\end{figure}

\begin{figure}
\includegraphics[scale=0.6]{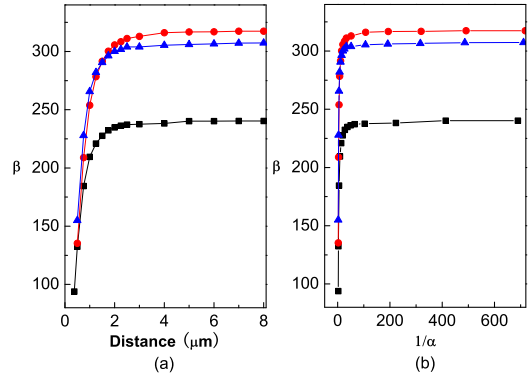}
\caption{\label{fig_7} (Color online) (a) The field enhancement factor($\beta$)
versus the spacing distance of the SWCNTs (\textit{d}); (b) $\beta$ versus
$1/\alpha$, for three sets of parameters: (1) squares:
$F_{\textrm{appl}}$=12.0\textrm{ V/$\mu$m} and \textit{L}=0.75\textrm{ $\mu$m},
(2) triangles: $F_{\textrm{appl}}$=10.0\textrm{ V/$\mu$m} and
\textit{L}=1.00\textrm{ $\mu$m}, (3) circles: $F_{\textrm{appl}}$=12.0\textrm{
V/$\mu$m} and \textit{L}=1.00\textrm{ $\mu$m}.}
\end{figure}

The reduction of field enhancement by screening can be seen more clearly in
Fig. \ref{fig_7} for a few sets of applied external fields and tube lengths.
Figure \ref{fig_7}(a) shows that for large spacing distance, where the
screening is week, the field enhancement factor tends to a constant. For a
given \textit{L}, the classical field enhancement factor should be a constant.
However, as one seen in \ref{fig_7}(b), that would be true only if the
screening could be ignored. For the instances of the \ref{fig_7}(b), the field
enhancement factor decreases sharply as $\alpha$ increase when $1/\alpha<20$.

%\begin{figure} %并排插入两个子图形
%\centering \subfigure[]{
%\label{fig_11_a} %%标记第一个子图形
%\includegraphics[scale=0.5]{fig_11_a}}
%\hspace{0.1in} \subfigure[]{
%\label{fig_11_b} %%标记第二个子图形
%\includegraphics[scale=0.5]{fig_11_b}}
%\caption{(a) The field enhancement factor($\beta$) versus \textit{L}; (b)
%$\beta$ versus $1/\alpha$, for three sets of parameters: (1) squares:
%$F_{\textrm{appl}}$=12.0\textrm{ V/$\mu$m} and \textit{L}=0.75\textrm{ $\mu$m},
%(2) triangles: $F_{\textrm{appl}}$=10.0\textrm{ V/$\mu$m} and
%\textit{L}=1.00\textrm{ $\mu$m}, (3) circles: $F_{\textrm{appl}}$=12.0\textrm{
%V/$\mu$m} and \textit{L}=1.00\textrm{ $\mu$m}.}
%\label{fig_11} %%标记整个图形
%\end{figure}

\section{\label{sec:level} Conclusions}

We have studied the field electron emission properties of arrays of
single-walled carbon nanotubes in the quantum level. As the spacing distance of
the SWCNTs exceeds a certain value (about the length of the tube), the emission
current density increases rapidly. When the spacing distance is large enough,
the current density decreases with the spacing distance as a parabolic
function. An optimal spacing distance that corresponds to the maximum current
density could be two to three times of the nanotube length. The exact value of
the ratio corresponding to the peak current density would depend on the applied
external field and the tube structure. That would be one of the reasons for the
discrepancy of experimental results of different groups. The screening effect
can be described by the factor ($\alpha$) defined in Eq. \ref{eq_3}, which is
found to be an exponential decreasing function of the ratio of the spacing
distance to the length of the SWCNTs, and is almost independent of the applied
external field. Therefore, the screening factor introduced here reflects an
intrinsic character of the array. When $\alpha<0.05$, the screening effect can
be ignored and each SWCNT of the array behaves as an individual emitter. For
large $\alpha$, on the other hand, the emission property of a SWCNT in the
array is obviously different from the individual SWCNT. The field enhancement
factor of the SWCNT array is not a linear function of the tube lengths as the
metal rod model predicted. And it depends on the applied external fields and
the spacing distances.  For a given tube length, it is a monotonically
decreasing function of the screening factor. Although the field enhancement
factors depends on the spacing distance (Fig. \ref{JvsL}(a)) in the same trend
as the classical calculation\cite{a9}, the quantitative discrepancy between
quantum simulation and classical calculation is large. To increase the emission
current density, one can either adjust the spacing distance or increase the
lengths of the tubes. For the (5,5) SWCNT array considered in the present
paper, we find that the current density is very small for the spacing distance
0.5 $\mu$m. For larger spacing distance (0.75 and 1.0$\mu$m, for instances),
the current density increases rapidly as the length increases until the length
is as large as 1.5 times of the spacing distance. For longer length, the
current density only increases slowly. It implies that lengths of SWCNTs of the
arrays need not be too long.
\newline

\textbf{Acknowledgement.} The project is supported by the National Natural
Science Foundation of China (Grant No. 10674182, 90103028, 90306016) and
National Basic Research Program of China (973 program 2007CB935500).


\newpage


\begin{thebibliography}{99}
\bibitem{a1} L. Nilsson, O. Groening, C. Emmenegger, O. Kuettel, E. Schaller, L. Schlapbach, H. Kind, J-M. Bonard, and K. Kern, Appl. Phys. Lett. \textbf{76}, 2071 (2000).
\bibitem{a2} J. S. Suh, K. S. Jeong, J. S. Lee, and I. Han, Appl. Phys. Lett. \textbf{80}, 2392 (2002).
\bibitem{a3} S. H. Jo, Y. Tu, Z. P. Huang, D. L. Carnahan, D. Z. Wang, and Z. F. Ren, Appl. Phys. Lett. \textbf{82}, 3520 (2003).
\bibitem{a4} Y. M. Wong, W. P. Kang, J. L. Davidson, B. K. Choi, W. Hofmeister, and J.H. Huang, Diamond \& Related Materials \textbf{14}, 2078 (2005).
\bibitem{a5} K. B. K. Teo, E. Minoux, L. Hudanski, F. Peauger, J-P. Schnell, L. Gangloff, P. Legagneux, D. Dieumegard, G. A. J. Amaratunga, and W. I. Milne, Nature \textbf{437}, 968 (2005).
\bibitem{a6} X. Q. Wang, M. Wang, H. L. Ge, Q. Chena, and Y. B. Xu, Physica E \textbf{30}, 101 (2005).
\bibitem{a7} X. Q. Wang, M. Wang, Z. H. Li, Y. B. Xu, and P. M. He, Ultramicroscopy \textbf{102}, 181 (2005).
\bibitem{a8} D. Nicolaescu, V. Filip, G. H. Takaoka, Y. Gotoh, and J. Ishikawa, J. Vac. Sci. Technol. B \textbf{25}, 472 (2007).
\bibitem{a9} G. S. Bocharov and A.V. Eletskii, Technical Physics \textbf{50}, 944 (2005).
\bibitem{a10} X. Zheng, G. H. Chen, Z. Li, S. Deng, and N. Xu, Phys. Rev. Lett. \textbf{92}, 106803 (2004).
\bibitem{a11} J. Peng, Z. Li, C. He, S. Deng, N. Xu, X. Zheng, and G. H. Chen, Phys. Rev. B \textbf{72}, 235106 (2005).
\bibitem{Dewar} M. J. S. Dewar and W. Thiel, J. Am. Chem. Soc, \textbf{99}, 4899 (1977).
\bibitem{a12} Rakwoo Chang, Gary S. Ayton, and Gregory A. Voth, J. Chem. Phys. \textbf{122}, 244716 (2005).
\bibitem{a13} A. M. Rao, E. Richter, S. Bandow, B. Chase, P. C. Eklund, K. A. Williams, S. Fang, K. R. Subbaswamy, M. Menon, A. Thess, R. E. Smalley, G. Dresselhaus, and M. S. Dresselhaus, Science \textbf{275}, 187 (1997).
\bibitem{a14} X. Blas$\acute{e}$, L. X. Benedict, E. L. Shirley, and S. G. Louie, Phys. Rev. Lett. \textbf{72}, 1878 (1994).
\bibitem{a15} R. A. Jishi, J. L. Bragin, and L. Lou, Phys. Rev. B \textbf{59}, 9862 (1999).
\bibitem{a16} Z. Li and W. Wang, Chin. Phys. Lett. \textbf{23}, 1618 (2006).
\end{thebibliography}
\end{document}